\documentclass[5p,twocolumn]{elsarticle}

\usepackage{lineno,hyperref}
\usepackage{amsmath,amsfonts}
\usepackage{algorithmic}
\usepackage{algorithm}
\usepackage{array}
\usepackage[caption=false,font=normalsize,labelfont=sf,textfont=sf]{subfig}
\usepackage{textcomp}
\usepackage{stfloats}
\usepackage{url}
\usepackage{verbatim}
\usepackage{graphicx}

\DeclareMathOperator\erfc{erfc}

\modulolinenumbers[5]

\journal{Journal of Optics and Laser Technology}
\begin{document}

\begin{frontmatter}

\title{Optical link acquisition for the LISA mission with in-field pointing architecture}

\author{Gerald Hechenblaikner, Simon Delchambre, and Tobias Ziegler}
\address{Airbus Defence and Space, 88039 Friedrichshafen, Germany}

\begin{abstract}
We present a comprehensive simulation of the spatial acquisition of optical links for the LISA mission in the in-field pointing architecture, where a fast pointing mirror is used to move the field-of-view of the optical transceiver, which was studied as an alternative scheme to the baselined telescope pointing architecture. The simulation includes a representative model of the far-field intensity distribution and the beam detection process using a realistic detector model, and a model of the expected platform jitter for two alternative control modes with different associated jitter spectra. 
For optimally adjusted detector settings and accounting for the actual far-field beam profile, we investigate the dependency of acquisition performance on the jitter spectrum and the track-width of the search spiral, while scan speed and detector integration time are varied over several orders of magnitude. 
Results show a strong dependency of the probability for acquisition failure on the width of the auto-correlation function of the jitter spectrum, which we compare to predictions of analytical models. Depending on the choice of scan speed, three different regimes may be entered which differ in failure probability by several orders of magnitude. We then use these results to optimize the acquisition architecture for the given jitter spectra with respect to failure rate and overall duration, concluding that the full constellation could be acquired on average in less than one minute.
Our method and findings can be applied to any other space mission using a fine-steering mirror for link acquisition.
\end{abstract}


\end{frontmatter}

\section{Introduction}
LISA is a science mission, led by ESA in partnership with NASA, aiming to detect gravitational waves in space in the frequency range from [0.1 mHz, 1 Hz] using optical interferometry\cite{danzmann2003lisa}. 
In order to obtain the high strain sensitivity to accomplish its goal, LISA performs interferometry in between 3 spacecraft (SC) in equilateral triangular configuration (see Fig.~\ref{fig_1}a) across on arm-length of 2.5 million km, where relative distance changes must be measured with an accuracy of around $10\,pm/\mathrm{\sqrt{Hz}}$. For each arm of the constellation, two one-way links are established between the two SC, which are used for heterodyne interferometry, ranging, data transmission and comparison of the reference clocks\cite{heinzel2011auxiliary}. Data products are transmitted to ground via classical RF links, where major noise sources, such as laser frequency noise\cite{muratore2020revisitation}, are removed to the extent possible in post-processing by a method referred to as Time-Delay-Interferometry (TDI)\cite{tinto2021time}.\\
However, before the optical links can be used in nominal science operations, they have to be established first during the acquisition phase. Owing to the large distance in between SC and the correspondingly large telescope apertures, the divergence angle of the transmitted beam is very small. Therefore, the large initial pointing uncertainty of the beam transmitted by one SC with respect to the actual position of its counterpart across the constellation arm prevent direct establishment of a link. Instead, it is required to perform an initial spiral search scan in the uncertainty plane (see Fig.~\ref{fig_1}b), where one of the two SC (SC1) is scanning the beam while the other SC (SC2) is waiting to receive the transmitted beam on its acquisition sensor. Once SC2 detects the scanning beam with sufficiently high signal-to-noise ratio (SNR) during the short time when the beam sweeps over its location, it may accurately point its beam into the direction of SC1. SC1 then likewise detects the beam transmitted by SC2 and re-orients itself in the corresponding direction. At this point, spatial acquisition of the 2 links of one constellation arm has been accomplished. This is then followed by the frequency acquisition phase, where the laser of SC2 is frequency offset-locked to the received laser from SC1 to enable coherent detection and heterodyne interferometry on the respective arm.\\
Link acquisition with a scanning beam is also used for TAIJI\cite{luo2020taiji, Gao2021laser}, a planned Chinese gravitational wave observatory, and for other missions performing interferometry \cite{wuchenich2014laser,gruber2014next} as  well as in most optical communication missions (see e.g.\cite{benzi2016optical}). A comprehensive review of all aspects of optical communication in space, including acquisition, is given in \cite{kaushal2016optical}. 

\subsection{Acquisition models}
Over the past two decades a number of papers have analyzed different aspects of optical link acquisition. Different search scan patterns were studied in \cite{scheinfeild2000} and the impact of beam divergence angle on scan time in \cite{hindman2004}. One of the biggest problems for acquisition is jitter of the scanning beam which may come from the platform accommodating the optical transceiver and/or a beam steering mechanism within the transceiver. This has the effect that the scanning beam of SC1 may miss SC2 while scanning the uncertainty region, resulting in acquisition failure. First simulations on this topic were also performed in \cite{hindman2004}, which investigated the impact of beam divergence and root-mean-square (RMS) jitter on the probability of acquisition which rises with increasing overlap between beams on adjacent tracks (see Fig.~\ref{fig_1}b) and falls with increasing RMS jitter. 
In \cite{friederichs2016vibration} a complex analytical model was used to assess acquisition failure under the influence of beam jitter. %which was considered in radial as well as in tangential direction. 
However, this model contained some crucial simplifications which impede its practical application. For example, it considered the probability of detection only for a specific SC position and did not average probabilities over the uncertainty distribution of SC positions. Furthermore, it assumed unlimited white Gaussian noise for the jitter spectrum, which results in an infinitely narrow correlation function. Consequently, jitter on the time-scale of the integration windows becomes fully uncorrelated, irrespective of the scan speed, which greatly increases the acquisition probability (as detection probabilities associated with the individual windows become largely independent).
Another analytical model \cite{ma2021satellite} related acquisition failure to geometric areas that are not covered by the jittering scanning beam. This model did not take into account that, for a reasonable beam overlap, any spacecraft missed on one track has a high probability of being detected after one spiral revolution on the adjacent track.\\
In \cite{hechenblaikner2021analysis} we derived an analytical model that relates the probability for acquisition failure to the beam radius, track width and RMS  beam jitter. The model accounts for potential detection on adjacent tracks and averages the probabilities over all possible positions, making predictions significantly more accurate than those of previous models that do not account for these effects.\\
However, all these models have in common that the effects of correlations are completely neglected and only RMS values of the amplitudes are used for predicting acquisition probabilities. 
Therefore, in an extension to our previous analysis, we studied how the spectral distribution and cut-off frequency affect the correlation between integration windows and scanning tracks and found that they strongly impact the acquisition probability\cite{hechenblaikner2022impact}. In this paper, we focus on one of the key parameters that may generally be adjusted by the acquisition architect, namely the scan speed, and show it can be tuned to enter 3 different regimes which differ greatly in acquisition probability and duration. On the one hand, scan speeds below a critical limit, which depends on the attitude jitter correlation time and the beam detection radius, increase acquisition probability, as individual integration windows become uncorrelated and integrated energies increase. On the other hand, scan speeds above another critical limit, which depends on correlation time and width of the uncertainty distribution, also increase acquisition probability, as jitter excursions on adjacent tracks become correlated.
Depending on the performance of the beam steering mechanism, bandwidth (BW) of the attitude control system, properties of the detector and the beam intensity profile, one or the other regime may be entered and the performance optimized through the system design. In this paper we present a full end-to-end model of spatial acquisition that accounts for all currently known effects and includes all relevant system entities, from attitude control to detector subsystem, in order to simulate acquisition probability. Results are found to agree very well with analytical models and predictions that are based on properties of the auto-correlation of the jitter spectrum.
%%%%%%%%%%%%%%%%%%%%%%%%%%%%%%%%%%%%%%%%%%%%%%%%%%%%%%%%%%%
\section{Mission architecture}
%%%%%%%%%%%%%%%%%%%%%%%%%%%%%%%%
\subsection{Optical design for in-field pointing}
\label{sec:opt_design}
\begin{figure}[!t]
\centering
\includegraphics[width=3.5in]{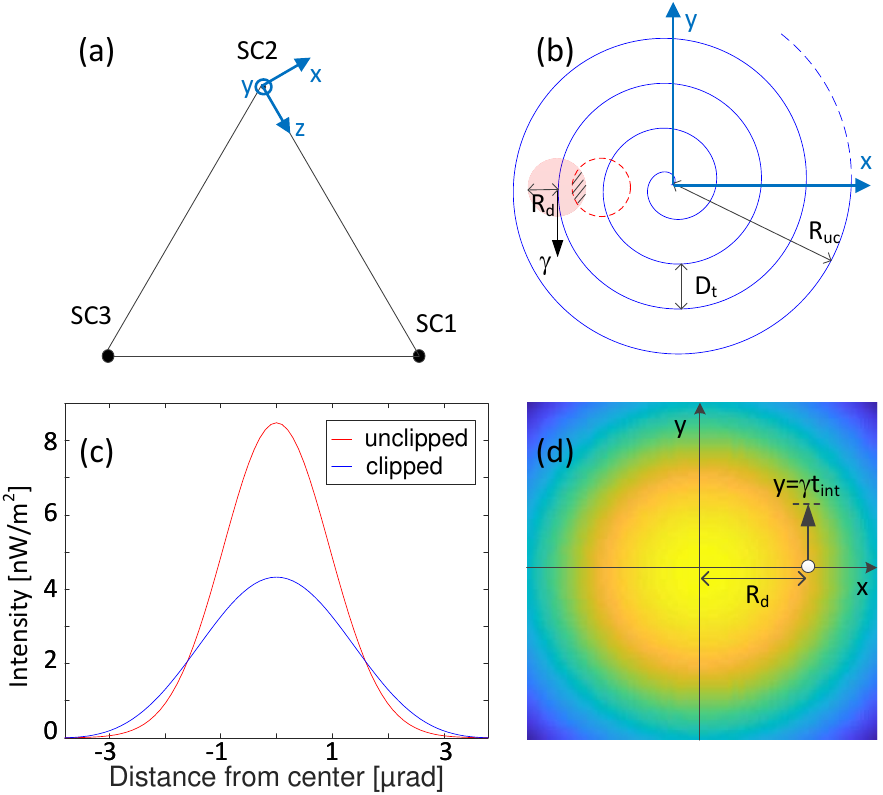}
\caption{(a) LISA constellation of 3 SC in equilateral triangular configuration; local coordinate system of SC2 given by blue arrows with y-axis out-of-plane. (b) Beam of radius $R_d$ scans the uncertainty plane with speed $\gamma$ along an Archimedean spiral of track width $D_t$ up to the maximum scan radius $R_{uc}$. Black hatched area indicates beam overlap between adjacent tracks.(c) Far-field intensities of the actual beam clipped by the system aperture (blue) and, for comparison, a virtual unclipped beam (red). (d) Definition of $R_d$: As the beam sweeps over SC2 (white circle) at a distance $R_d$, its intensity distribution is integrated over a range given by the product of scan speed $\gamma$ and integration time $t_{int}$.\label{fig_1}}
\end{figure}

The LISA mission is currently undergoing a Phase B1 study by competing industrial teams under the lead of ESA. 
The system parameters used for the subsequent analysis are based on the published mission architecture at the conclusion of the preceding ``Mission Assessment Phase'' which is summarized in\cite{danzmann2011LISA}.
The three SC form an equilateral triangle with an arm-length of 5 million km (see Fig.~\ref{fig_1}a). Each SC is equipped with two optical transceivers to establish links to both its neighbours in the constellation. The transceiver telescopes have an aperture size of 40 cm and support a field-of-view (FoV) of $400\,\mathrm{\mu rad}$ diameter\cite{danzmann2011LISA,weise2009alternative}. The laser has a wavelength of $1064\,\mathrm{nm}$ and operates with a minimum output power of $\mathrm{2\,W}$\cite{danzmann2011LISA}. We also assume all the transmission losses for the transmitting and receiving path specified in \cite{danzmann2011LISA}, but we consider a higher efficiency for delivery of the laser power to the optical bench (80\%).\\
Choosing the ratio of telescope aperture radius $R_A$ to Gaussian beam waist $w_0$ to be 1.12, we maximize the peak power received by the remote telescope in the far-field, as beam truncation and diffraction losses are balanced in an optimal way. In that case $\approx 8\%$ of power is clipped by the system aperture and the far-field intensity pattern is almost Gaussian, as shown by the blue line in Fig.~\ref{fig_1}c, but slightly broader in the center and dropping-off more sharply towards the tails. We obtain a $1/e^2$ beam radius of $2.5\,\mathrm{\mu rad}$ ($12.5\,\mathrm{km}$)  and a link loss factor of $4.5\times 10^{-10}$ across the 5 million km distance between SC.\\
%%%
%%%
If we account for an estimated loss of 10\% due to wavefront-errors and mode impurities, the received power is approximately $540\,\mathrm{pW}$. Considering transmission losses through the receiver optics as given in \cite{danzmann2011LISA} and noting that only a fraction of light is used for acquisition sensing (and most of it for interferometry), we assume that $22\,\mathrm{pW}$ are received by the acquisition sensor under ideal alignment, i.e. if the receiving SC2 is located in the center of the transmitted beam.\\
As the orbit of the LISA constellation is not actively stabilized but left to evolve freely, distances between SC change and there is a seasonal variation of the constellation angle which deviates by up to $\pm 1$ deg from the nominal 60 deg of an equilateral triangle, requiring continuous re-adjustments of the transceiver pointing. While there are several options to accomplish this, a decision was made to implement the ``telescope pointing'' architecture for LISA at the end of the Mission Formulation Phase\cite{danzmann2011LISA}, where the complete transceiver assembly, also referred to as MOSA, is actuated.\\
An alternative scheme, referred to as ``In-Field Pointing" (IFP), has been proposed in\cite{weise2009alternative}. This scheme overcomes some of the challenges of telescope-pointing, but this comes at the price of a more complex optical design. In the IFP architecture, a steering mirror is placed in the intermediate pupil plane of the telescope. Over the past years feasibility of this concept has matured by developments of the required mechanism \cite{witvoet2015realization} and experimental demonstration of relevant performances\cite{hasselmann2021lisa,hasselmann2022elimination}. The beam can be moved quasi continuously and the expected jitter due to the driving electronics is expected to be $<0.1\,\mathrm{\mu rad}$ RMS on sky which is much smaller than the platform jitter and therefore neglected in the following analysis. 
As the transmitted beam must only move along one direction to compensate changes of the constellation angle, the original IFP scheme only actuates the steering mirror in 1 degree-of-freedom (dof). However, this can be generalized to actuate 2 dof, as commonly found for most missions employing optical links in space (see e.g. \cite{wuchenich2014laser,benzi2016optical,hindman2004}), to which the same approach as outlined in this paper may be applied. In the case of LISA, implementing IFP in 2 dof may increase complexity of the optical design and degrade tilt-to-length coupling performance, but the very high achievable scan rates greatly improve acquisition performance, as discussed further down.

%%%%%%%%%%%%%%%%%%%%%%%%
\subsection{Spacecraft control}
\label{sec:SC_control}
\begin{figure}[!t]
\centering
\includegraphics[width=3.5in]{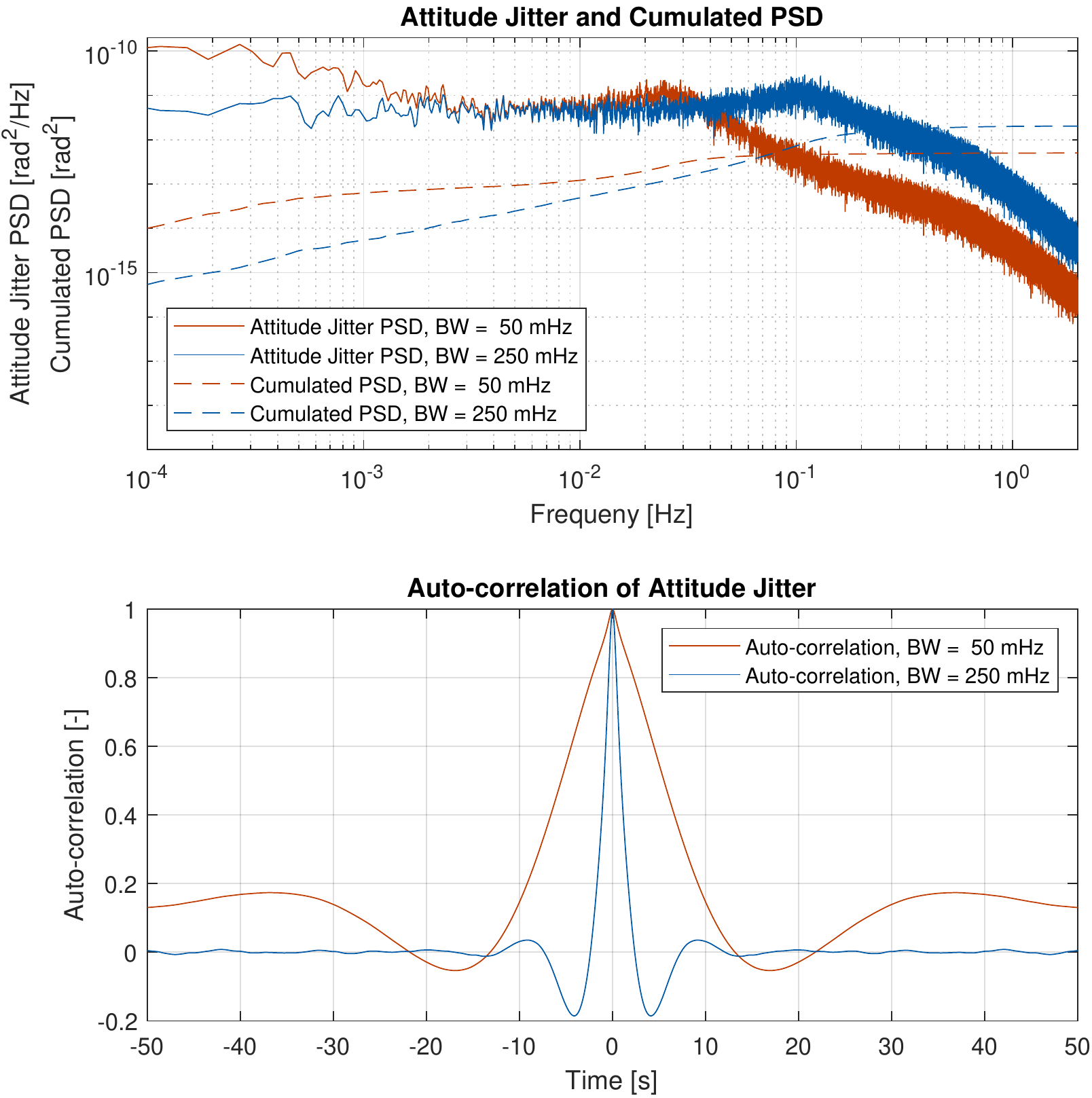}
\caption{(Top) Power spectral densities and cumulative integral of SC jitter for $50\,\mathrm{mHz}$ (red) and $250\,\mathrm{mHz}$ (blue) control bandwidth. (Bottom) Auto-correlation functions associated with the jitter time series.\label{fig_2}}
\end{figure}

The proposed attitude control system controls the inertial attitude direction of the laser beam, combining platform closed-loop inertial control with open-loop relative control of the IFP mirror. As the latter has a negligible control error, the inertial attitude jitter is driven by the platform closed-loop control, which employs a micro-propulsion thruster actuation system and star-trackers (STR) for inertial reference guidance.\\
For the ultra-low thrust noise of the micro-propulsion system we assume a spectral density of $1\,\mathrm{\mu N/\sqrt{Hz}}$, as demonstrated for the cold-gas system of LISA Pathfinder\cite{schleicher2018orbit} that is also used in LISA, which translates into a torque noise of $1\,\mathrm{\mu Nm/\sqrt{Hz}}$ through the equivalent lever arm of 1~m. 
The star-trackers are assumed to have a total cross-boresight error of 1 arcsec ($\mathrm{1\sigma}$) at a rate of $10\,\mathrm{Hz}$.
The plant $G$ of the control system is modeled as a double integrator with a moment of inertia $J$ and a Laplace transform given by: $G(s)=1/(Js^2)$.
The decoupled attitude controller $K$ is a PD-controller with a phase-lead filter and a second order low-pass to provide sufficient phase margin and suppression of measurement noise. We consider a low- and mid-frequency control BW of $50\,\mathrm{mHz}$ and $250\,\mathrm{mHz}$, respectively:
\begin{align*}
K_{low}(s)&=&\frac{309 s^2+508.4 s+87.1}{s^3+39.9 s^2+786.4 s+2750}\\
K_{mid}(s)&=&\frac{1202 s^2+2651 s+1317}{s^3+39.9 s^2+786.4 s+2750}
\end{align*}
In steady-state, the closed-loop attitude jitter of the inertial beam direction is driven by the micro-propulsion and star-tracker noise. For lower bandwidths, the control-loop is more sensitive to disturbance noise, whereas for increasing bandwidths measurement noise drives the attitude performance. 
This is also observed in Fig.~\ref{fig_2}~(top), where the inertial attitude power spectral densities (PSD) are illustrated for the two controllers. The results are derived from single-input, single-output time series simulations, assuming perfect mass property knowledge and unit alignments. 
The cumulated integral of the power spectral density is plotted in Fig.~\ref{fig_2}~(top) as well, where the total average power or variance of the attitude
jitter can be determined from the asymptotic value. The RMS jitter $\sigma_n$ increases from $0.7$ to $1.4\,\mathrm{\mu rad}$ (per degree-of-freedom) when the bandwidth is increased from $50\,\mathrm{mHz}$ to $250\,\mathrm{mHz}$, which clearly indicates that star-tracker noise is the driving noise source. 
Figure~\ref{fig_2}~(bottom) plots the auto-correlation functions of the jitter time series which have a peak-to-zero width ($\tau_0$) of $13.2\,\mathrm{s}$ and $2.4\,\mathrm{s}$, and a half-maximum width ($\tau_{1/2}$) of $5.5\,\mathrm{s}$ and $0.9\,\mathrm{s}$ for $50\,\mathrm{mHz}$ and $250\,\mathrm{mHz}$ BW, respectively. Note that these widths essentially determine the ``correlation properties'' of the jitter which greatly impact the probability of acquisition failure, as we will discuss in section~\ref{sec:scan_rate}.
%%%%%%%%%%%%%%%%%%%%%%%%%%%%%%%%%%%%%%%%%%%%%%%%%%%%%%%%%%%%%%%%%%%%
\section{Acquisition Performance}
%%%%%%%%%%%%%%%%%%%%%%%%%%%%%%%%%%%%%
\subsection{Acquisition Requirements}
\label{sec:acquisition_requirements}
Accurate pointing of the beam transmitted from SC1, which must also account for the point-ahead-angle, is required in order to hit SC2. However, there is an uncertainty with respect to the correct pointing direction for which the major contributors are (1) misalignments of the transceiver axis with respect to its nominal direction in the STR reference frame (primarily caused by ground-to-orbit effects), (2) orbit determination errors, and (3) inertial attitude knowledge error. Assuming that the misalignment (1) is calibrated in-orbit\cite{cirillo2009control} with a residual knowledge error of $5.1\,\mathrm{\mu rad}$ ($1\sigma$) and that the contribution from ephemeris errors is given by $12.5\,\mathrm{\mu rad}$  \cite{ESA2017CDF} ($1\mathrm{\sigma}$), the root-sum-square of the total uncertainty is found to be $\sigma_{uc}=14.5\,\mathrm{\mu rad}$. 
The parameter $\sigma^2_{uc}$ describes the variance of the Gaussian uncertainty distributions for angles in- and out-of the constellation plane which can be expressed as a Rayleigh distribution for the probability density function (PDF) in radial direction of the uncertainty plane\cite{scheinfeild2000}:
\begin{equation}
PDF(r)=\frac{r}{\sigma_{uc}^2}e^{-\frac{r^2}{2\sigma_{uc}^2}}
\end{equation}
As the beam divergence angle is much smaller than $\sigma_{uc}$, the beam will generally not hit SC2 if SC1 aims to point it in the direction where SC2 is believed to be. Therefore, SC1 must perform a spiral scan to cover the uncertainty region of radius $R_{uc}$ (see Fig.~\ref{fig_1}b) which extends up to $R_{uc}=3.44\sigma_{uc}$ for $3\sigma$ ($99.73\%$) confidence to locate the waiting SC2 within it, or up to $R_{uc}=4.40\,\sigma_{uc}$ for $4\sigma$ confidence. Note that the respective boundaries of $3.44\sigma_{uc}$ and $4.40\sigma_{uc}$ refer to the integration limits of the Rayleigh distribution which differ from those of a Gaussian distribution.\\
Once SC2 is hit by the beam from SC1, it must re-orient its own beam into the direction of the received beam with sufficient accuracy to hit SC1. This requires a sufficiently high SNR of the beam image on its matrix detector to achieve the necessary centroiding accuracy. 
The total of centroiding error $\sigma_p$ and other sources of error, such as jitter and drifts, must not exceed the effective beam radius (approximately $3\,\mathrm{\mu rad}$), otherwise the beam will miss SC1. Allocating $3\sigma_p = 0.9\,\mathrm{\mu rad}$ provides sufficient margin for the other contributors. Considering that the pixel-resolution on sky is $0.94\,\mathrm{\mu rad}$ (see next section), the RMS centroiding error must remain below $\approx 1/3$ of a pixel, which is achievable for an SNR of 40 and a CoG centroiding algorithm (see e.g. \cite{thomas2006comparison,gao2020high}). 
%%%%%%%%%%%%%%%%%%%%%%%%%%%%%%%%%%%%%
\subsection{Detection performance for variable scan rates}
\begin{figure}[!t]
\centering
\includegraphics[width=3.5in]{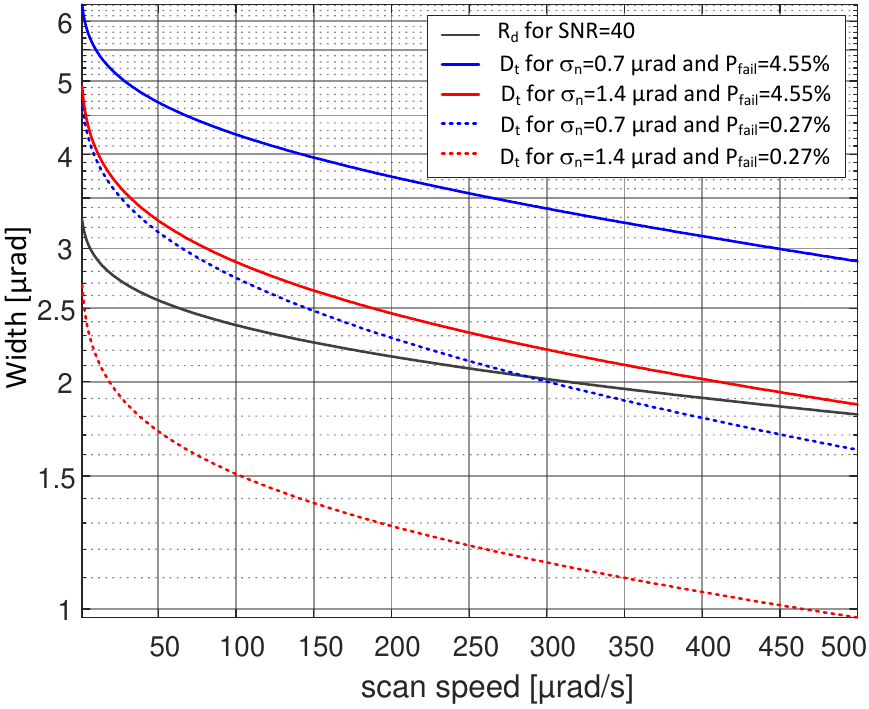}
\caption{Radius of detection $R_d$ (black curve) and associated track widths $D_t$ (colored curves) plotted against scan speed $\gamma$. The track widths denoted by the blue and red curves were obtained for RMS beam jitter $\sigma_n$ of \text{$0.7\,\mathrm{\mu rad}$} ($50\,\mathrm{mHz}$ BW) and $1.4\,\mathrm{\mu rad}$ ($250\,\mathrm{mHz}$ BW), respectively. Solid and dashed lines indicate track-widths which result in $4.55\%$ and $0.27\%$ probability of acquisition failure, respectively.\label{fig_3}}
\end{figure}

Assuming a cumulative signal $S_{sig}$ in units of photons incident on a centroiding region of $N_{pix}$ pixels during an integration time $t_{int}$, one can express the standard equation for SNR of the acquisition sensor (see e.g. \cite{hindman2004}) as follows:
\begin{equation}
SNR=\frac{S_{sig}Q_e}{\sqrt{S_{sig}Q_e+N_{pix}t_{int}(N_r^2f_s+S_{sl}Q_e+D_{dark})}}
\label{eq:SNR}
\end{equation}
Here, $Q_e$ is the quantum efficiency of the detector, $S_{sl}$ [photons/pixel/s] the background stray light flux , $D_{dark}$ [$e^-$/pixel/s] the detector dark current, and $N_r$ [$e^-$/pixel] the readout noise at a frame rate of $f_s$ frames per second.\\
We now want to evaluate $S_{sig}$ for the specific conditions of the scanning beam path relative to SC2, for which we introduce the radius of detection $R_d$, as explained in \cite{hechenblaikner2022impact}. It defines the maximal angular distance the scanning beam may have (at the point of closest approach) when sweeping past SC2 in order to be detected with sufficient SNR for the given scan speed $\gamma$ and integration time $t_{int}$. Considering that $S_{sig}$ is obtained by integrating the power incident on the detector during the integration period and dividing by the photon-energy $h\nu$, we obtain the following relation:
\begin{equation}
S_{sig}=\frac{I_0 A_{ap}T_l}{h\nu}\int_0^{t_{int}}\!\!\!\! F(R_d,\gamma t)dt=\frac{P_{max}}{h\nu\,\gamma}\int_0^{\gamma t_{int}}\!\!\!\! F(R_d,y)dy,
\label{eq:Signal}
\end{equation}
where $I=I_0F(x,y)$ describes the far-field intensity distribution with a peak intensity $I_0$, $x$ and $y$ refer to the angles in the uncertainty plane (see Fig.~\ref{fig_1}a,b) with the coordinate origin placed in the center of the beam at the start of integration, and the beam is assumed to sweep along the y-axis past SC2 located at $x=R_d$, as shown in Fig.~\ref{fig_1}d.
The product of the collecting aperture $A_{ap}$, the transmission losses from the aperture to the detector $T_l$, and the peak intensity $I_0$ can be combined to yield the maximum power incident on the detector when SC2 is located in the beam center: $P_{max}=22\,\mathrm{pW}$ (see section \ref{sec:opt_design}).\\ 
In order to obtain exemplary numbers for the LISA in-field pointing architecture, we choose the TekWin SH640 CMOS camera that has been studied in the context of the TAIJI mission in \cite{gao2020high,Gao2021laser}, where essential detector parameters are listed. We assume an off-axis telescope with a magnification of $M=134$, as used for recent telescope pointing architecture designs \cite{sanz2018telescope}. Stray light intensities are very sensitive to the telescope design.  From the scatter ratios for in-field-pointing telescope designs 1 and 2 presented in \cite{livas2017elisa}, we estimated radiant intensities of $12\,\mathrm{\mu W/sr}$ and $200\,\mathrm{\mu W/sr}$, respectively, where an attenuation factor of $1/40$ (due to power-splitters) between the last telescope mirror and the acquisition sensor is considered. Therefore, in the following analysis we conservatively assume a stray light radiant intensity of $I_{rad}=200\,\mathrm{\mu W/sr}$ incident on the detector.
The point-spread-function (PSF) of the telescope (on sky) is assumed to be $3.75\,\mathrm{\mu rad}$ (FWHM on sky), i.e. close to the diffraction limit, and sampled with a pixel resolution of $0.94\,\mathrm{\mu rad}$ (on sky) within a region-of-interest (ROI) that is between 2 to 3 times larger than the PSF. The stray light power received by each pixel within its field-of-view $\Omega=1.6\times 10^{-8}\,\mathrm{sr}$ is given by $I_{rad}\Omega=3.2\,\mathrm{pW}$. The associated shot noise is the dominant noise source and approximately 10 times larger than readout-noise or dark current. The detector is assumed to read-out the data with a nominal rate of $f_s=300\,\mathrm{Hz}$ (full frame) and up to $500\,\mathrm{Hz}$ for sub-frames imaging the acquisition FoV.\\    
Inserting Eq.~\ref{eq:Signal} into Eq.~\ref{eq:SNR} we can determine the value for $R_d$ that yields an SNR of 40, which is easily performed by a combination of inversion of Eq.~\ref{eq:SNR} and iterative numerical integration of Eq.~\ref{eq:Signal}. We obtain $R_d=3.3\,\mathrm{\mu rad}$ for a scan speed of $\gamma=1\,\mathrm{\mu rad/s}$ and an integration time of $t_{int}=1\,\mathrm{s}$. The value of $R_d$ is found to be insensitive to small increases or decreases of $t_{int}$ with an optimum given by $\gamma t_{int}\approx 1\,\mathrm{\mu rad}$, i.e. approximately $1/3\,R_d$. For this reason, when increasing $\gamma$ in our computations, we decreased $t_{int}$ in reverse proportion in order to preserve optimal conditions for detection. The resulting radius of detection $R_d$ is found to decrease by almost a factor 2 from 3.3 to \text{$1.8\,\mathrm{\mu rad}$} when increasing the scan speed from 1 to $500\,\mathrm{\mu rad/s}$, as shown by the solid black curve in Fig.~\ref{fig_3}.
%%%%%%%%%%%%%%%%%

%%%%%%%%%%%%%%%%%%%%%%%%%%%%%%%%%%%%%%%%%
\subsection{Acquisition failure dependency on track-width}
\label{sec:analytical_model}
\begin{figure}[!t]
\centering
\includegraphics[width=3.5in]{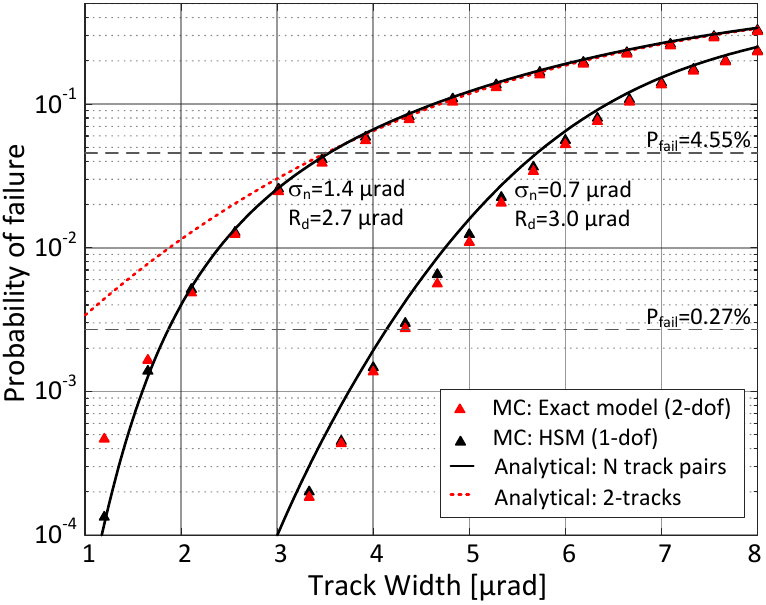}
\caption{Probability of acquisition failure $P_{fail}$ plotted against track-width. Analytical predictions are given by the solid lines, MC simulation results by the red and black triangles for the exact model and the HSM, respectively. Left curves computed for BW=$250\,\mathrm{mHz}$ ($\sigma_n=1.4\,\mathrm{\mu rad}$) and $\gamma=30\,\mathrm{\mu rad/s}$ ($R_d=2.7\,\mathrm{\mu rad}$). Right curves computed for BW=$50\,\mathrm{mHz}$ ($\sigma_n=0.7\,\mathrm{\mu rad}$) and $\gamma=6\,\mathrm{\mu rad/s}$ ($R_d=3.0\,\mathrm{\mu rad}$). MC simulations per data point: $6\times 10^4$.\label{fig_4}}
\end{figure}

We would now like to investigate the dependency of acquisition failure on the track width of the search spiral. We have previously derived an analytical model in \cite{hechenblaikner2021analysis} that relates the probability of acquisition failure $P_{fail}$ to the RMS value of the beam jitter $\sigma_n$, the radius of detection $R_d$, and the track width $D_t$: 
\begin{equation}
P_{fail}=\frac{1}{2D_t}\int_0^{\frac{D_t}{2}}\! dx_t\text{erfc}\left(\frac{R_d-x_t}{\sqrt{2}\sigma_n}\right)\!\text{erfc}\left(\frac{x_t-D_t+R_d}{\sqrt{2}\sigma_n}\right).
\label{eq:P_fail}
\end{equation}
It is important to note that this model considers the probability that if the scanning beam misses SC2 on one spiral track, it may still ``hit'' it on the adjacent track. Decreasing the track width $D_t$ increases the beam overlap $OV=2R_d-D_t$ between adjacent tracks (see Fig.~\ref{fig_1}b), which greatly reduces $P_{fail}$.
Large overlap $OV$ and strong jitter imply that $D_t\leq\sqrt{R_d^2+\sigma_n^2}$. In this case, one must consider that the scanning beam may hit SC2 on multiple adjacent tracks, which requires extending the simple ``2 track'' model of Eq.~\ref{eq:P_fail} to account for $N$ pairs of tracks\cite{hechenblaikner2022impact}. The probability of failure $P_{fail}$ is then found to be:
\begin{align}
P_{x}(x)&=\prod_{k=-N+1}^{N}\frac{1}{2}\left[\erfc\left(\frac{x+R_d-k\cdot D_t}{\sqrt{2}\sigma_n}\right) \right.\nonumber\\
&+\left.\erfc\left(\frac{R_d+k\cdot D_t-x}{\sqrt{2}\sigma_n}\right)\right]\nonumber\\
P_{fail}&=\frac{2}{D_t}\int_{0}^{D_t/2} dx P_{x}(x)\label{eq:P_tot},
\end{align}
where $P_{x}(x)$ denotes the probability of acquisition failure if SC2 is located at position $x$ in between one track ($x=0$) and half the distance to the adjacent track ($x=D_t/2$), and $P_{fail}$ is the average over $P_x$ to account for a random position of SC2.
The black solid lines of Fig.~\ref{fig_4} were obtained from Eq.~\ref{eq:P_tot} with $N=3$ for $250\,\mathrm{mHz}$ BW and $\gamma=30\,\mathrm{\mu rad/s}$ (left) and for $50\,\mathrm{mHz}$ BW and $\gamma=6\,\mathrm{\mu rad/s}$ (right), respectively. We find that $R_d=2.7\,\mathrm{\mu rad}$ and $R_d=3.0\,\mathrm{\mu rad}$ for the two scan speeds, respectively, which can be determined from the general dependency of $R_d$ on $\gamma$ represented by the black curve of Fig.~\ref{fig_3}. These values for $R_d$, as well as the RMS jitters associated with the respective BW have been used as input to the analytical models . While there is no difference between the prediction of Eqs.~\ref{eq:P_tot} and \ref{eq:P_fail} for the case of $50\,\mathrm{mHz}$ BW, the ``2-track'' model predictions for $250\,\mathrm{mHz}$ BW (red-dashed line) start deviating from the black line for $D_t<3\,\mathrm{\mu rad}$, which is expected due to the narrow $D_t$ and large $\sigma_n$.
Looking at the trend of the curves in Fig.~\ref{fig_4}, it is apparent that $P_{fail}$ is higher for the larger jitter (upper curve) and drops sharply towards smaller track-widths.\\ 
%%%%%%%%%%%
%%%%%%%%%%%%%
We also find very good agreement between the analytical predictions and the results of Monte Carlo (MC) simulations which take the exact SC jitter noise spectrum as input and accurately model the integration of the far-field intensity distribution by the detector, only counting a detected signal of $\mathrm{SNR}>40$ towards a successful acquisition. For a general description of our Monte Carlo approach see \cite{hechenblaikner2022impact}.
The red triangles in Fig.~\ref{fig_4} refer to simulation results obtained with the exact simulation model, while the blue triangles refer to results from a much simpler model, which we refer to as ``Hard Sphere Model'' (HSM). In the HSM only radial jitter is applied and an acquisition failure is counted, if the distance between the center of the scanning beam and the waiting SC2 exceeds $R_d$ for the entire scan. The good agreement between the HSM and the exact simulation suggest that (1) for the given parameters the complexity of the detection process is well approximated by a single parameter ($R_d$) and that (2) the tangential jitter has no notable impact on the failure rate. Note that both these assumptions have entered the derivation of the analytical models in Eqs.~\ref{eq:P_fail} and \ref{eq:P_tot} in order to remove unnecessary complexity.\\
The upper horizontal line in Fig.~\ref{fig_4} indicates that a probability of failure of $4.55\%$ is obtained if $D_t=3.5\,\mathrm{\mu rad}$ and $D_t=5.7\,\mathrm{\mu rad}$ for $250\,\mathrm{mHz}$ and $50\,\mathrm{mHz}$ BW, respectively. In a similar way, the track-widths associated with $4.55\%$ ($2\sigma$) and $0.27\%$ ($3\sigma$) probability of acquisition failure are determined for a large range of scan speeds and plotted in Fig.~\ref{fig_3} as solid and dashed colored lines, respectively. 
For example, if the track-width is chosen according to the blue solid line of Fig.~\ref{fig_3} for $50\,\mathrm{mHz}$ BW, $P_{fail}$  is expected to remain constant at a value of $4.55\%$ for all scan speeds. However, as we will discuss in the next section, this is not the case due to the emergence of correlation effects which are not considered in the analytical model.

%%%%%%%%%%%%%%%%%%%%%%%%%%%%%%%%%%
\subsection{ Acquisition failure dependency on scan rate}
\label{sec:scan_rate}
The analytical model of section~\ref{sec:analytical_model} assumes that a jitter excursion remains quasi constant while the beam passes by SC2, which implies that the beam transit time $T_{pass}=2R_d/\gamma$ must be much smaller than the width of the jitter auto-correlation function: $\tau_{1/2}\gg 2R_d/\gamma$. If, on the other hand, the scan speed $\gamma$ is reduced and the correlation time becomes smaller than $T_{pass}$, the individual integration windows (in our case $\approx 6$) become mutually uncorrelated, which reduces $P_{fail}$ \cite{hechenblaikner2022impact}.
Therefore, we expect $P_{fail}$ to decrease continuously towards lower scan speeds, as also the auto-correlation of the beam jitter evaluated at the transit time $r_{xx}(T_{pass})$ decreases with $\gamma$. Based on the auto-correlation width at half maximum $\tau_{1/2}$, we specify a guide value $\gamma_{-}$ of the scan speed, where the impact of this effect on $P_{fail}$ should be clearly recognizable:
\begin{equation}
\label{eq:gamma_low}
\gamma_{-}=\frac{2R_d}{\tau_{1/2}},
\end{equation}
Another effect comes into play when scan speeds are so large that jitter fluctuations are positively correlated between two spiral revolutions, which also reduces $P_{fail}$. Correlations exist, if the peak-to-zero width $\tau_0$ of the auto-correlation function exceeds the mean spiral revolution time $T_{rev}=2\pi R_m/\gamma$, where the mean spiral radius $R_m$ is found from the Rayleigh distribution for the uncertainty to be $R_m=\sqrt{\pi/2}\sigma_{uc}$. In this case, a large excursion of the beam from its nominal path onto the adjacent track has the effect that it misses SC2, but -because the excursion persists for the next revolution due to a positive correlation- the beam is very likely to ``hit'' SC2 while moving along the adjacent track\cite{hechenblaikner2022impact}.
The critical scan speed $\gamma_{+}$ above which this effect occurs is found from the inequality $\tau_0 > T_{rev}$ to be:
\begin{equation}
\gamma_{+}=\frac{2\pi\sqrt{\pi/2}\sigma_{uc}}{\tau_{0}}.
\label{eq:gamma_high}
\end{equation}
The two critical frequencies, $\gamma_{-}$ and $\gamma_{+}$, define three regimes which can be entered in any acquisition architecture, depending on the choice of scan speed $\gamma$, as listed in Table~\ref{table_1}. Note that the two correlation effects discussed above are weak but not negligible in regime 2 so that the peak value of $P_{fail}$ is expected to be slightly lower than the prediction $P_{lim}$ of the analytical model of Eq.~\ref{eq:P_tot}. Using the widths of the auto-correlation functions given in section~\ref{sec:SC_control}, we obtain for $(\gamma_{-},\gamma_{+})$ the values of $(1.1,8.7)\,\mathrm{\mu rad/s}$ and $(7.4,47.6)\,\mathrm{\mu rad/s}$ for $50\,\mathrm{mHz}$ BW and $250\,\mathrm{mHz}$ BW, respectively. 
\begin{table}
\begin{center}
\caption{Acquisition regimes in relation to scan speed $\gamma$}
\label{table_1}
\begin{tabular}{ccl}
Regime 1: & $\gamma<\gamma_{-}$ & $P_{fail}$ increases with $\gamma$  \\
\hline
Regime 2: & $\gamma_{-}<\gamma<\gamma_{+}$ & $P_{fail}$ peaks \\
\hline
Regime 3: & $\gamma_{+}<\gamma$ & $P_{fail}$ decreases with $\gamma$
\end{tabular}
\end{center}
\end{table}

\begin{figure}[!t]
\centering
\includegraphics[width=3.5in]{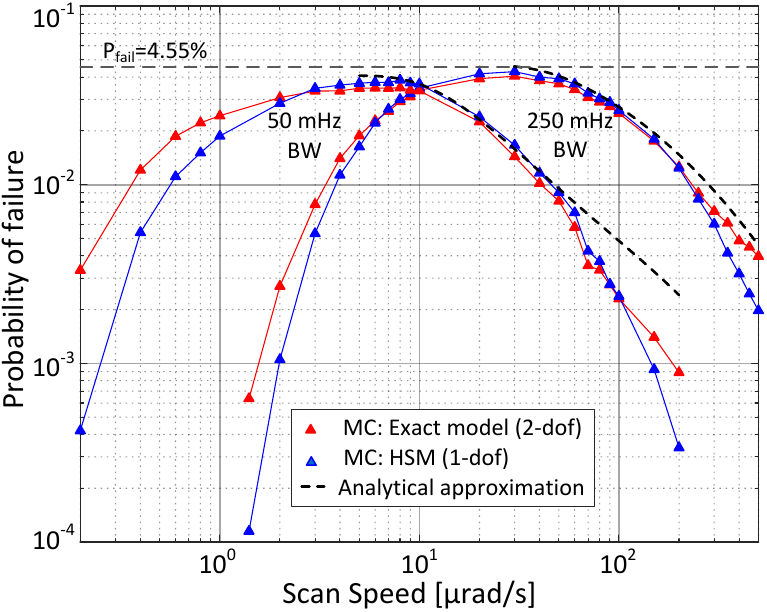}
\caption{Probability of acquisition failure $P_{fail}$ plotted against scan speed for BW of $50\,\mathrm{mHz}$ (left curves) and $250\,\mathrm{mHz}$ (right curves). MC simulation results are given by red (exact model) and blue (HSM) triangles. Analytical approximations are given by black dashed lines. MC simulations per data point: $4\times 10^4$ to $2\times 10^5$.\label{fig_5}}
\end{figure}

We performed MC simulations where the scan speed is varied between $0.2$ and $500\,\mathrm{\mu rad/s}$, which implies that $R_d$ is changing as shown by the black curve of Fig.~\ref{fig_3}. The track-width $D_t$ was chosen such that $P_{fail}$ predicted by the analytical model remained constant at $P_{lim}=4.55\%$, which required adjusting $D_t$ to the values given by the blue ($50\,\mathrm{mHz}$ BW) and red ($250\,\mathrm{mHz}$ BW) solid lines of Fig.~\ref{fig_3} for each specific scan speed. The simulation results are shown in Fig.~\ref{fig_5}.\\
As expected from the preceding discussion, $P_{fail}$ at first increases with scan speed and reaches a peak close to the limit $P_{lim}$ (horizontal dashed line) before decreasing again towards higher scan speeds. The peak is reached first for the configuration of $50\,\mathrm{mHz}$ BW, because its auto-correlation width is over 5 times larger than the one for $250\,\mathrm{mHz}$ BW. The deviation of the simulation data from the horizontal limit is entirely due to the effects of correlations which are not considered in models that only account for the RMS values of the jitter amplitude. For fast scan speeds and using a simplistic model, it is reasonable to expect that the simulated $P_{fail}$ scales with $P_{lim}$ and the auto-correlation function $r_{xx}(\tau)$ as $P_{fail}=P_{lim}[1-r_{xx}(T_{rev})]$, where $T_{rev}=2\pi r/\gamma$ is the spiral revolution time for a radial position $r$\cite{hechenblaikner2022impact}. Averaging this expression over the Rayleigh distribution of the positional uncertainty, we obtain the black dashed lines plotted in Fig.~\ref{fig_5} which are in reasonable agreement for lower scan speeds but diverge towards higher speeds.
It is apparent that extremely small failure probabilities can be achieved for very low or very high scan speeds. This obviously also affects the mean acquisition time which we discuss in the next section.

%%%%%%%%%%%%%%%%%%%%%%%%%%%%%%%%%%%%%%%%%%%%%%%%%%%%%%%
\subsection{The mean acquisition search time}
\begin{figure}[!t]
\centering
\includegraphics[width=3.5in]{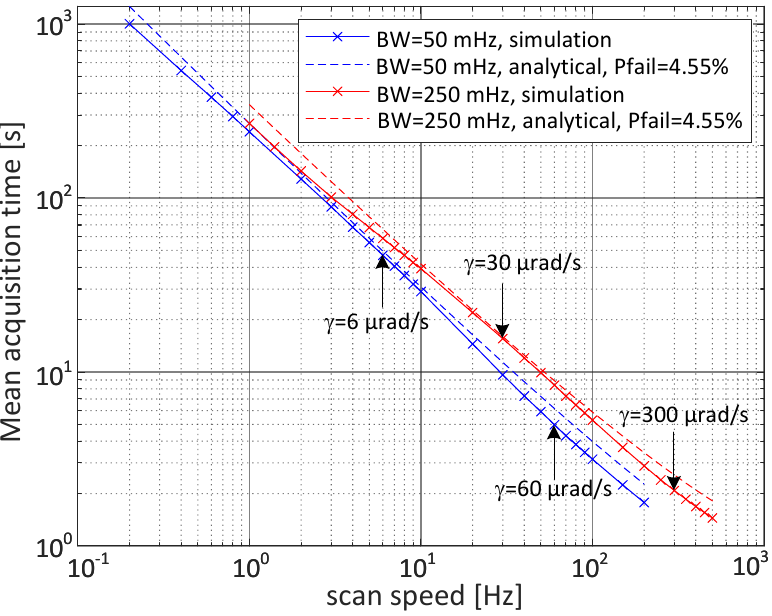}
\caption{Mean acquisition search time plotted against scan speed for the same MC simulations as shown in Fig.~\ref{fig_5} with a BW of $50\,\mathrm{mHz}$ (blue curve) and $250\,\mathrm{mHz}$ (red curve). Analytical predictions for an assumed $P_{fail}=4.55\%$ are given by the dashed curves, respectively. MC simulations per data point: $4\times 10^4$ to $2\times 10^5$.\label{fig_6}}
\end{figure}

If $P_{fail}$ is high for a given set of acquisition parameters, multiple scans may have to be performed to conclude the acquisition successfully. Before a scan can be repeated, the full uncertainty region of radius $R_{uc}$ must be scanned, which takes a time $T_{uc}=R_{uc}^2\pi/(D_t\gamma)$ to complete, while the mean acquisition search time of a single scan is given by $T_{1s}=2\pi\sigma_{uc}^2/(D_t\gamma)$\cite{li2011analytical, hechenblaikner2021analysis}. Using the expression for the mean search time of multiple scans $T_{ms}$ derived in \cite{li2011analytical} and inserting the above expressions for $T_{uc}$ and $T_{1s}$, we find after some rearrangements:
\begin{equation}
T_{ms}=T_{1s}\left(1+\frac{F^2}{2}\frac{P_{fail}}{1-P_{fail}}\right),
\label{eq:multi_scan}
\end{equation}
Here, the factor $F=3.44$ relates the width of the uncertainty distribution to the assumed maximum scan radius ($R_{uc}=F\sigma_{uc}$) and acts as a weighting factor for the impact of $P_{fail}$ on the mean search time of multiple scans.
If the scan speed $\gamma$ is increased within regime 3, this decreases $T_{1s}$ (scaling $\propto 1/\gamma$) as well as $P_{fail}$, which both contribute to decreasing $T_{ms}$ according to equation~\ref{eq:multi_scan}. On the other hand, our assumed reduction of $D_t$ with $\gamma$, as given by the solid red and blue lines in Fig.~\ref{fig_3}, increases $T_{1s}$ and therefore $T_{ms}$, but to a much smaller degree so that the overall trend is a strong decrease of of $T_{ms}$ with $\gamma$.\\
Fig.~\ref{fig_6} plots the mean acquisition search time $T_{ms}$ against scan speed $\gamma$ over a range of over 3 orders of magnitude for the same simulation data as plotted in Fig.~\ref{fig_5}. The simulation results for both configurations, $250\,\mathrm{mHz}$ BW (red solid line) and $50\,\mathrm{mHz}$ BW (blue solid line) closely approach the respective analytical predictions (dashed lines) in the range of scan speeds defined by ``regime 2'' in Table \ref{table_1} but fall below them in the other regimes. This is because the track width $D_t$ for the simulation performed at each scan speed was chosen such that the analytical model of \ref{eq:P_tot} predicted a constant failure probability of $P_{fail}=4.55\%$ (see red and blue solid lines of Fig.~\ref{fig_3}). However, as we have seen in the discussion of Fig.~\ref{fig_5}, the actual failure probabilities outside ``regime 2'' can be much lower due to the effect of correlation, which leads to a reduced mean acquisition time according to Eq.\ref{eq:multi_scan}

\begin{figure}[!t]
\centering
\includegraphics[width=3.5in]{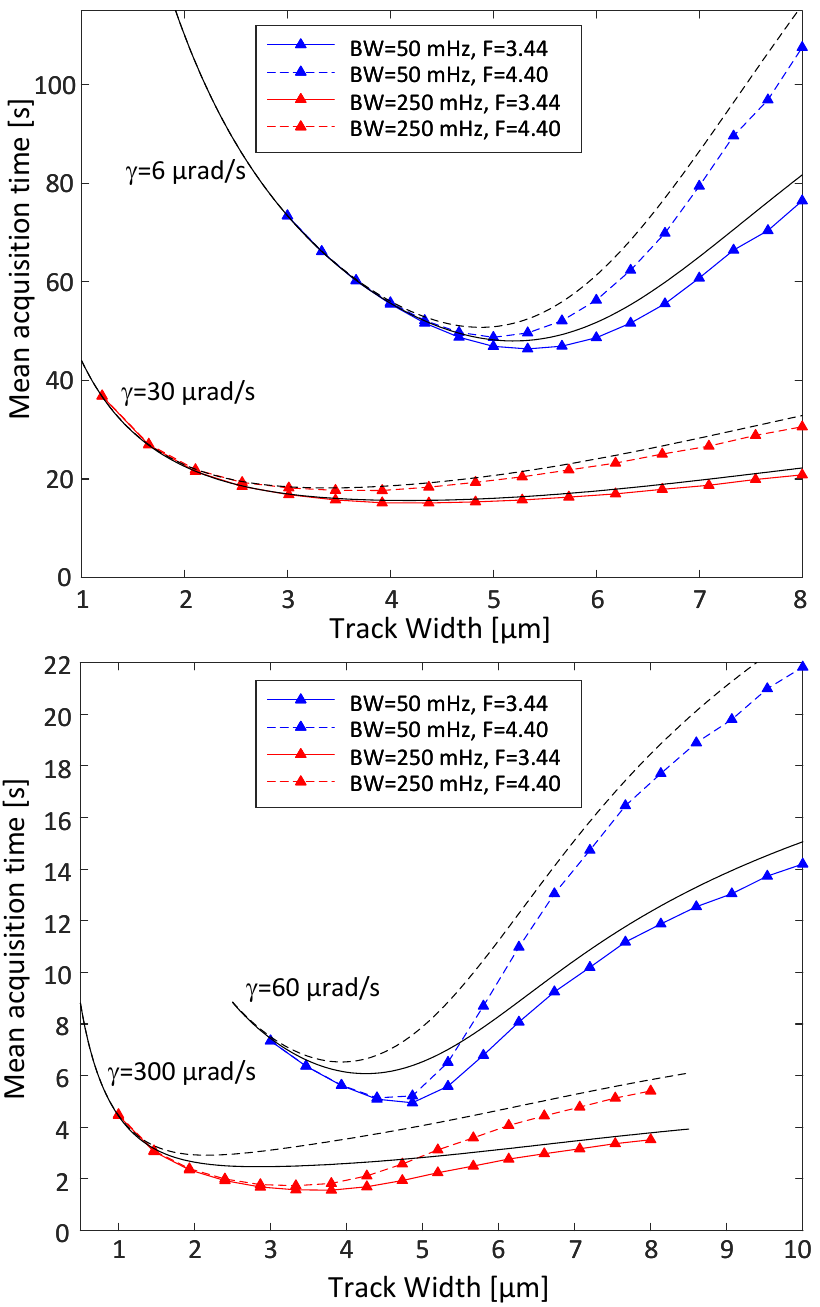}
\caption{Mean acquisition search time plotted against track width for simulations with a BW of $50\,\mathrm{mHz}$ (blue curves) and $250\,\mathrm{mHz}$ (red curves). Analytical predictions are given by the black lines. The factor $F$, relating the maximum scan radius $R_{uc}$ to the standard deviation of the uncertainty distribution $\sigma_{uc}$, is given by $F=3.44$ for the solid lines and $F=4.40$ for the dashed lines, respectively. (Top) Plots are for scan speeds $\gamma$ of $6\,\mathrm{\mu rad/s}$ and $30\,\mathrm{\mu rad/s}$ for the two bandwidths, respectively.
(Bottom) Plots are for scan speeds $\gamma$ of $60\,\mathrm{\mu rad/s}$ and $300\,\mathrm{\mu rad/s}$ for the two bandwidths, respectively. MC simulations per data point: $6\times 10^4$.\label{fig_7}}
\end{figure}

One should note that the mean acquisition search times $T_{ms}$ plotted in Fig.~\ref{fig_6} generally do not correspond to the optimum configuration which is determined by the combination of track width $D_t$ and resulting failure probability $P_{fail}$ that minimizes the mean acquisition search time of Eq.~\ref{eq:multi_scan} for a given scan speed $\gamma$. The blue lines in Fig~\ref{fig_7} (top) plot $T_{ms}$ against track width $D_t$ for $\gamma=6\,\mathrm{\mu rad/s}$ and a BW of $50\,\mathrm{mHz}$, the red lines for $\gamma=30\,\mathrm{\mu rad/s}$ and a BW of $250\,\mathrm{mHz}$, corresponding to the same parameters as in Fig.~\ref{fig_4}. The solid and dashed lines were derived for different sizes of the scanning region, given by $F=3.44$ and $F=4.40$, respectively. It is apparent that $T_{ms}$ for $F=4.40$ increases more rapidly towards larger track widths than $T_{ms}$ for $F=3.44$, because the increasing failure probabilities incur a higher loss of time when repeating the scan over a larger scanning region, as can also be seen in Eq.~\ref{eq:multi_scan}. The simulation data are in good agreement with the analytical predictions obtained from Eqs.~\ref{eq:P_tot} and \ref{eq:multi_scan}, because the simulation parameters are within regime 2 of Table~\ref{table_1}, where correlations of the beam jitter in between tracks are small and the analytical model is valid. For the smaller scanning region ($F=3.44$) and the BW of $50\,\mathrm{mHz}$  we find that the optimum acquisition search time of 48.0~s is found for a track width of $D_t=5.2\,\mathrm{\mu rad}$ and an associated failure probability of $P_{fail}=2.16\%$. This is close to the time 49.5~s which we obtain for the track width of $D_t=5.7\,\mathrm{\mu rad}$ associated with our default failure probability of $P_{fail}=4.55\%$. 

If the scan speed is greatly increased, we move from regime 2 into regime 3 of Table~\ref{table_1}, where correlations between tracks become large and strongly suppress the failure probabilities. In this case, the results of MC simulations (colored curves) yield significantly shorter acquisition search times than the analytical predictions (black curves), as can be seen in Fig.~\ref{fig_7} (bottom). In particular, this is also the regime entered by optical communication missions, where jitter correlations have not been accounted for in previous publications despite their significant impact on the acquisition performance.

%%%%%%%%%%%%%%%%%%%%%%%%%%%%%%%%%%%%%%%%%%%%%%%%%%%%%%%
\subsection{Acquiring the constellation}
From Fig.~\ref{fig_6} we find that by increasing the scan speed from $1\,\mathrm{\mu rad/s}$ to $200\,\mathrm{\mu rad/s}$ for a BW of $50\,\mathrm{mHz}$, the mean acquisition search time decreases from 240~s~s to 1.8~s.
Increasing the scan speed beyond $200\,\mathrm{\mu rad/s}$ does not gain much, because the overall acquisition time, which is obtained from the search time plus additional (quasi constant) time delays in the acquisition sequence, is dominated by the long travel time of light between the two SC. It is given by $T_{light}=16.7\,\mathrm{s}$ and enters the sequence at three distinct occasions: (1) for the scanning beam to hit the waiting SC2, (2) for the beam emitted by the SC2 to hit the scanning SC1 after adjusting its pointing, and (3) for the beam emitted by SC2 to hit SC1 once more after adjusting its pointing. On top of this, we add a flat time of $T_{rd}=1\,\mathrm{s}$ to account for the reorientation of the IFP mirror into the direction of the received beam on two occasions.
The mean acquisition of a single arm $T_{1arm}$ is then given by the sum of a ``constant offset time'' $T_{const}=3T_{light}+2T_{rd}=52.1\,\mathrm{s}$ and the mean search time $T_{ms}$, which yields $T_{1arm}=T_{const}+T_{ms}=53.9\,\mathrm{s}$ for $\gamma=200\,\mathrm{\mu rad/s}$ and a BW of $50\,\mathrm{mHz}$.\\
If the three arms are acquired sequentially, the mean time to acquire the constellation is simply three times larger: $T_{3arm}=3\,T_{1arm}$. However, the IFP architecture supports parallel acquisition of all three arms simultaneously. Considering the extremely low failure probability we have for large scan speeds ($P_{fail}\approx 10^{-3})$), we find that $T_{ms}\approx T_{1s}$, meaning that only a single scan is needed. The mean time to conclude the single scans for 3 arms in parallel $T_{3ps}$ can be computed from the PDF given in \cite{hechenblaikner2021analysis}, and we find after some transformations
\begin{equation}
T_{3ps}=\int_{0}^{\infty}dt\,\frac{3t}{T_{1s}}e^{-\frac{t}{T_{1s}}}\left(1-e^{-\frac{t}{T_{1s}}}\right)^2=H(3)T_{1s},
\label{eq:3parallel}
\end{equation}
where $H(3)=1.83$ is the harmonic function evaluated for the argument 3. 
Using Eq.~\ref{eq:3parallel} we obtain for the mean acquisition time in a parallel scheme $T_{3arm}=T_{const}+T_{3ps}=55.4\,\mathrm{s}$, i.e. we expect to acquire all 3 arms of the LISA constellation in less than 1 minute.

\section{Conclusion}
We have presented a detailed model for the spatial acquisition of optical links in the LISA mission, using the alternative in-field pointing architecture.
Starting from essential requirements, we discussed and analyzed the full chain and inter-dependencies of acquisition parameters, including SNR, scan speed $\gamma$, integration time $t_{int}$, radius of detection $R_d$, track width $D_t$, as well as RMS amplitude $\sigma_n$ and auto-correlation $r_{xx}(\tau)$ of the beam jitter.
We studied the impact on acquisition performance for two simulated SC control models of different bandwidth and jitter properties.
Unlike previous publications we did not only consider a simple ``beam divergence angle'' to describe the acquisition but computed the exact radius of detection $R_d$ to achieve the required SNR over a large range of scan speeds, accounting for a representative beam intensity distribution as well as realistic detector performance parameters and noise sources.\\ 
We then used the calculated values of $R_d$ as input to a previously derived analytical model to calculate the track-width required to fall below a certain probability of acquisition failure  $P_{fail}$ and obtained very good agreement to the results of representative MC simulations. However, we found that this agreement only holds in a certain range of scan speeds, referred to as ``regime 2'' in Table \ref{table_1}, where correlations of beam jitter do not show up. 
Outside this regime, when scan speeds are either very low (regime 1) or very high (regime 3), MC simulations showed that correlations may reduce $P_{fail}$ by several orders of magnitude. This paper is the first to introduce and define the boundaries between ``3 acquisition regimes'' and accurately simulate the acquisition failure for scan speeds in a range spanning over 3 orders of magnitude. The 3 regimes can be successively entered by increasing the scan speed, which is a concept applicable to any mission scenario and not specific to the one for LISA covered here.
We then simulated the mean acquisition search time for a single link, accounting for the possibility that multiple scans may be needed in case of failure, and compared it to analytical predictions, showing how the minimum time for a given scan speed can be obtained through the right choice of track width and associated failure probability .
Finally, we determined the mean constellation acquisition time for an optimized parallel acquisition architecture and found that the full LISA constellation of 3 SC can be acquired in less than 1 minute, which is two orders of magnitude faster than what was presented in\cite{cirillo2009control}.

%%%%%%%%%%%%%%%%%%%%%%%%%%%%%%%%%%%%%%%%%%%%%%
\section*{Acknowledgments}
The authors are grateful to A. Sell for detailed information on the IFP mechanism and control electronics and acknowledge stimulating discussions with T. Lamour, C. Greve, and R. Gerndt.

%%%%% References %%%%%

\end{document}